\documentclass[twocolumn,showpacs,pra,aps]{revtex4-1}

\usepackage{amsmath}
\usepackage{amsfonts}
\usepackage{amssymb}

\usepackage{graphicx}
\usepackage{natbib}

\usepackage {graphicx}\title{⤢}

\newcommand{\ds}{\displaystyle}

\newcommand{\be}{\begin{equation}}
\newcommand{\ee}{\end{equation}}
\newcommand{\beq}{\begin{eqnarray}}
\newcommand{\eeq}{\end{eqnarray}}

\newcommand{\W}{\Omega}

\newcommand{\bnn}{\begin{eqnarray*}}
\newcommand{\enn}{\end{eqnarray*}}

\newcommand{\ket}{\rangle}

\begin{document}
\title{Creation of quantum entangled states of Rydberg atoms via chirped adiabatic passage}

\author{Elliot Pachniak$^1$}

\author{Svetlana A. Malinovskaya$^1$ }\email[Corresponding author.\\ {\em E-mail address:}]{ smalinov@stevens.edu} 

\affiliation{$^1$Department of Physics, Stevens Institute of Technology, Hoboken, NJ 07030, USA}

\date{\today}

\begin{abstract}
Entangled states are crucial for modern quantum enabled technology which makes their creation key for future developments. In this paper, a robust  quantum control methodology is presented to create entangled states of two typical classes, the W and the Greenberger-Horne-Zeilinger (GHZ). It was developed from the analysis of a chain of alkali atoms $^{87}Rb $ interaction with laser pulses, which leads to the two-photon transitions from the ground to the Rydberg states with a predetermined magnetic quantum number. The methodology is based on the mechanism of two-photon excitation, adiabatic for the GHZ and non-adiabatic for the W state, induced by overlapping chirped pulses and governed by the Rabi frequency, the one-photon detuning, and the strength of the Rydberg-Rydberg interactions. \end{abstract}

\maketitle
Trapped Rydberg atoms can be used as viable systems to study collective phenomena in many-body physics. The most prominent features of Rydberg atoms are strong long-range interatomic interactions and an extraordinary long lifetime of the Rydberg states \cite{Ga05,Sa10}. They make Rydberg atoms an effective platform to simulate interacting spin systems in order to understand and control quantum properties, e.g., magnetism, coherence and entanglement. Due to the importance of these properties, the ability to generate entangled states on demand is paramount to modern quantum-enabled technology. 
Typical, complementary classes of entangled states 
are the GHZ states~\cite{7,11,12}, important for quantum information processing and quantum metrology, and the W states~\cite{8,Dur01} relevant for quantum communication and quantum cryptography ~\cite{Ni00,4,5,13,14}. 
 There has been a number of proposals to generate the GHZ and the W entangled states with electron or nuclear spin systems in a variety of arrangements. In \cite{Ch17} such states are generated on nuclear spins by the global control method implying transverse magnetic fields  and using optimization procedure for the spin rotation and free evolution parameters. In \cite{LukinArc}  a programmable quantum simulator is used for a sophisticated manipulation of entanglement in Rydberg qubit   states of atomic arrays incorporating local effective detunings for higher selectivity of excitation by optimal field shapes. The GHZ state was generated using superpositions of the atomic ground state and a Rydberg state in two antiferromagnetic configurations in 20 atoms individually trapped  in one-dimensional array. The Rabi frequency was less than the interaction strength of two Rydberg atoms on neighboring sites resulting in the Rydberg blockade forbidding the excitation of the adjacent atoms to the Rydberg state. Optimal control methods were used to find laser pulses that maximize the GHZ state preparation fidelity. In \cite{Gu18,Li19} a theory of  the stimulated Raman adiabatic passage scheme (STIRAP) combined with the Rydberg blockade was presented to create GHZ states (but not the W states) in the manifold of low-lying, metastable states of atomic ensemble. The Rydberg blockade created by the control atom excited to the Rydberg state eliminated losses from the transitional Rydberg states of atomic ensemble when STIRAP pulses were applied.   In contrast to previous proposals presented, in this paper a  method is developed to operate on the selective (hyper)fine magnetic sublevels of the ground and the Rydberg atoms using $\mu  s $ pulses having MHz chirp rate. Quantum control of population transfer by the pulse chirping leads to a desired superposition state of either the GHZ or W type.   The distinguishing feature of our quantum control method for creating multipartite entangled states is the use of a simple analytical function for the phase of the incident pulses to perform two-photon transitions. Such phase may be robustly produced by a liquid crystal pulse shaper. The Rabi frequencies associated with the two incident pulses have to completely overlap as this condition is known to aid to two-photon adiabatic passage \cite{Ma07,Ma072}. 
  The methodology is applicable to a system of N trapped atoms. The limitation on the value of N  originates from the resolution of the energy levels of the collective states whose gap becomes smaller with  increasing number of atoms. As the density of the energy levels rises, avoided crossings in the field-interaction picture become spectrally close, which may cause multiple couplings between dressed states. For the GHZ state formation, the limitation is also determined by the time of the chirp turn-off to avoid further undesirable crossings of the ground state with transitional states. A conceptual demonstration of the proposed methodology is presented  in the framework of a linear chain of three individually trapped atoms.

The W and the GHZ states are generated in a model of a chain of alkali atoms $^{87}Rb$ trapped in an optical lattice with  periodic structure. Each atom is considered  a three-dimensional subsystem with the ground state $|g\rangle$, the transitional, excited state $|e\rangle$ and the Rydberg state $|r\rangle$. 
The GHZ state is a quantum superposition of all subsystems in the ground state $|g\rangle$ and all in an excited state, such as $|r\rangle$. For a three-atomic system,  the GHZ state reads
\be
|GHZ\ket = \ds{|rrr\ket + |ggg\ket\over\sqrt{2}}.
\ee
The W state is a quantum superposition of all possible pure states in which one subsystem is in state $| g\rangle$, while all other ones are in state $|r\rangle$. For the three-atomic system, the W state reads
 \be \label{Wdefine}
 |W\ket = \ds{|grr\ket+|rgr\ket + |rrg\ket\over\sqrt{3}}.
 \ee
 The energies of these states depend on the pairwise interaction strength between Rydberg atoms $V_{ij}$ determined by the location of Rydberg atoms in the optical lattice, which specifies the distance between them $r_{ij}$. The  interaction strength is proportional to $r_{ij}^{-6}$ and $r_{ij}^{-3} $ for the van der Waals and dipole-dipole type of interactions respectively. 

\begin{figure}
	\hspace{-2.5cm}	
	\includegraphics[width=9cm]{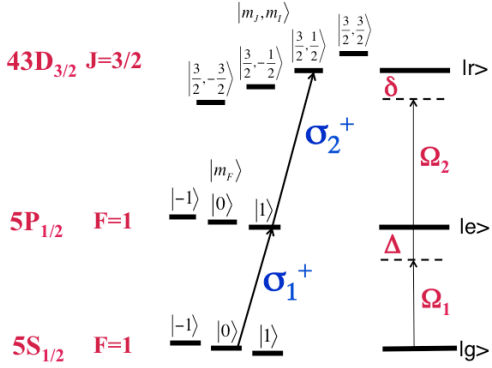}
	\caption{ \small{The manifold of magnetic sublevels in ultracold $^{87}Rb$, relevant for the studies of the electron dynamics.}}
	\label{fig1}
\end{figure}

 The repulsive van der Waals interaction between atom pairs is  $V_{ij}=C_6/r^6_{ij}$, where $C_6$ is the van der Waals interaction constant.  In a linear one-dimensional optical lattice of length s, n atoms are equidistantly separated by a lattice spacing $a=s/(n-1)$ giving the next neighbor interaction equal to $V_{ij}=C_6/(s/(n-1))^6$. Since $V_{ij}$ is distance dependent, it may be used as a control parameter to prepare the target state having a predetermined energy by manipulating the optical lattice. 

In order to generate an entangled state with predetermined properties, we investigated the mechanisms of two-photon excitation of a chain of coupled three-level atoms and revealed the range of parameters of the laser fields which induce the desirable transitions. These transitions stir the atomic system into a superposition state with predetermined entanglement properties. 
This is the essence of a quantum control  scheme design. For generation of the W and GHZ states, the control scheme makes use of two linearly chirped laser fields. The Rabi frequency, the one-photon detuning and the strength of Rydberg-Rydberg interaction are the key elements in the quantum control scheme, with a specific correlation of values between them leading to a desirable superposition state.

In alkali atoms having one valence electron, the spin information is conveniently encoded in the magnetic sublevels existing within the hyperfine states ($m_F$). For the Rydberg atoms, spin states are the projections of 
the electron spin-angular momentum on the quantization axis ($m_J$). In Fig.(\ref{fig1}), a schematic of addressed magnetic subevels - the spin states -  is shown. These are the ground $|g\rangle$ state $|5S_{1/2}, F=1, m_F=0\rangle$, the intermediate $|e\rangle$ state $|5P_{1/2}, F=1, m_F=1\rangle$ and the Rydberg $|r\rangle$ state $|43D_{3/2}, m_J=3/2, m_I=1/2\rangle$ \cite{Jo08}; they form a three-level ladder subsystem. The Zeeman splitting is made to exceed the collective state energy shift induced by Rydberg blockade to address magnetic sublevels selectively. Here we choose the next-neighbor Rydberg-Rydberg interaction strength to be $ 60 MHz$ and  the value of $B=10^2 G$. In $^{87}Rb$ such magnetic field induces the Zeeman split of the $ |5S_{1/2},F=1 \rangle$ state equal to $\Delta E_{|Fm_F\rangle}=-70 MHz$ and that of the $ |5P_{1/2},F=1 \rangle$ state equal to $\Delta E_{|Fm_F\rangle}=-23 MHz$. The Rydberg state $|Jm_j, I m_I\rangle$ gets the Zeeman split $\Delta E_{|J m_J,I m_I\rangle} = 158 MHz$ for $J=3/2, g_J =1.13$. This splitting is within the range of the fine structure, which is a few hundreds MHz $(\sim 1/n^3)$ for the Rydberg state of n=43 \cite{Ga05}. Thus all three chosen magnetic sublevels are within a respective, single (hyper)fine splitting, securing no overlap of different (hyper)fine states.  These magnetic sublevels are coupled by two $\sigma^+_{1,2}$ circularly polarized pulses having carrier frequencies  $\omega_1(t)$ and $\omega_2(t)$. The pulse duration is 1 $\mu s$, which satisfies the condition $1/\tau  < \Delta E_{Zeeman}$ to resolve Zeeman structure.

A chain of three-level ladder subsystems, coupled via the $|r\rangle$ states to reproduce the van der Waals interactions, is used as a model system to describe atoms in a one-dimensional, periodic, optical lattice and to design the W and the GHZ  spin entangled states. The total Hamiltonian that describes the interaction with optical fields reads
$\hat H_{tot} = \hat H_0 + \hat V_\W.$   
 Here $\hat H_0$ is the atomic Hamiltonian, which includes the Rydberg-Rydberg interaction between atoms, and the 
 $\hat V_\W$ is the atom-field interaction Hamiltonian, which describes the interaction of optical pulses with trapped atoms. \\
 
  {\bf A methodology to create the W and the GHZ states.} The quantum control methodology to create the entangled, multipartite spin states  is deduced from the dressed state analysis of $\hat H_0 + \hat V_\W$, which  
for N coupled three-level ladder subsystems  
reads \cite{Ma17}
\begin{eqnarray} \label{HAM_singleatom}
&\hat H(t)=\Sigma_{i=1}^{N} (\Delta - \alpha_1 (t-t_c) )  \sigma^{(i)}_{ee} + \nonumber \\ &\Sigma_{i=1}^{N} (\delta - (\alpha_1+\alpha_2)(t-t_c))  \sigma^{(i)}_{rr} + \nonumber \\
&\Sigma_{i=1}^{N} [\frac{\Omega_{01}(t)}{2}(\sigma^{(i)}_{ge} + \sigma^{(i)}_{eg}) + \frac{\Omega_{02}(t)}{2}(\sigma^{(i)}_{er} + \sigma^{(i)}_{re})] + \nonumber\\
&\Sigma_{i,j=1}^{N} V_{ij}  \sigma^{(i)}_{rr}\sigma^{(j)}_{rr}.
\end{eqnarray}
Here $\sigma^{(i)}_{km} = |k \rangle \langle m |$, where $k,m = g,e,r,$ the $V_{ij}$ describes the Rydberg-Rydberg interaction, the $\Delta$ and $\delta$ are the one-photon and  the two-photon detunings, the $\Omega_{01}(t)=\mu_{eg}E_{01}(t)/\hbar$ and the $\Omega_{02}(t)=\mu_{re}E_{02}(t)/\hbar$ are the time-dependent Rabi frequencies that couple the $|g\rangle \rightarrow  |e\rangle$ and $|e\rangle \rightarrow |r\rangle$ states respectively, shown in Fig.(\ref{fig1}), and $\alpha_{i},$ i=1,2  is the linear chirp rate of the applied pulses, which are 
\begin{equation}
E_{i}(t)=E_{0i}(t) \sin{(\omega_i (t-t_c)+ \alpha_i (t-t_c)^2 /2)}, \label{pulse}
\end{equation} 
having the pulse envelope  $ E_{0i}(t)= E_{0i} e^{-(t-t_c)^2/(2\tau_0^2)} $ with the peak value $E_{0i}$ at the central time $t_c$. 

For a three-atomic linear chain, this Hamiltonian was written in the field interaction representation in a collective state basis $| k, m, j \rangle$, ($k,m,j = g,e,r$), having dimension $3^N=27$. A truncated matrix Hamiltonian reads
\begin{tiny}
\begin{equation}
\hat H(t)=
\begin{bmatrix}
3 \omega_1& \Omega_{01}&0&\dots&0&\dots&0\\
\Omega_{01} & 2 \omega_1 +w_2 & \Omega_{02} &\dots&0&\dots&0\\
0 & \Omega_{02} & 2 \omega_1 + w_3 &\dots&0&\dots&0 \\
\vdots&\vdots&\vdots&\ddots&\vdots&\dots&0 \\
0&0&0&\dots&\omega_1 +2w_3+2 V_{23}&\dots&0 \\
\vdots&\vdots&\vdots&\vdots&\vdots&\ddots&\vdots \\
0&0&0&0&0&\dots&3w_3+2 V_{max} \\
\end{bmatrix} \label{threeatom} \nonumber
\end{equation}
\end{tiny}
The diagonal elements represent the energy levels of the bare collective states in the field interaction representation. Out of 27 collective state energies, 12 are unique and read
\begin{eqnarray}
E_1^{ggg}(t)=3\omega_1 \nonumber  \\
E_2^{geg,gge,egg}(t)=2\omega_1 + \omega_2 \nonumber \\
E_3^{gee,eeg,ege}(t)=\omega_1 + 2\omega_2 \nonumber \\
E_4^{eee} (t)=3\omega_2 \nonumber \\
E_5^{grg,rgg,ggr} (t)=2\omega_1 + \omega_3 \nonumber \\
E_6^{ger,erg,rge,gre,egr,reg}(t) =\omega_1 + \omega_2 \nonumber \\
E_7^{ere,eer,ree} (t) =2\omega_2 + \omega_3 \nonumber \\
E_8^{grr,rrg} (t) =\omega_1 + 2\omega_3+2V_{32} \nonumber \\
E_9^{rgr} (t) =\omega_1 + 2\omega_3+2V_{31} \nonumber \\
E_{10}^{err,rre} (t)  =\omega_2 + 2\omega_3+2V_{32} \nonumber \\
E_{11}^{rer} (t)  =\omega_2 + 2\omega_3+2V_{31} \nonumber \\
E_{12}^{rrr} (t) =3\omega_3+2V_{max}. \label{energy}
\end{eqnarray}
Here $\omega_1=0,$ $\omega_2 (t)  =\Delta - \alpha_1 (t-t_c)$ and $\omega_3 (t)  = \delta - (\alpha_1 + \alpha_2) (t - t_c)$, and $V_{max} = V_{21}+V_{32}+V_{31}$. In numerical calculations, the values of $V_{21}=V_{32}=V$ and two values of $V_{31}=V/2$ and $V_{31}=V/2^6$ giving $V_{max} = 5/2V$ and $V_{max} = 2 \frac{1}{64} V$ were used. Other parameters are $\alpha_1 = \alpha_2=\alpha$,  $\Omega_{01}=\Omega_{02}$ and the value of $\delta$ satisfying the condition $3 \delta = - 2V_{max} $ (for GHZ) and $ \delta = - V$ (for W) to compensate for the collective energy shift,  (see $E_{12}^{rrr} (t)$,   $E_{8}^{grr,rrg} (t)$ and $E_9^{rgr} (t)$  in Eq.(\ref{energy}) respectively).

 The time-dependent Schr\"odinger equation with this matrix Hamiltonian 
 was used for the numerical analysis of the evolution of the collective states which was solved numerically using Runge-Kutta method \cite{Kutta}.
 
  The quantum control scheme was deduced from the mathematical analysis of the Hamiltonian in the field-interaction representation. The energies of the required bare states were equated to derive the time of the resonance, (see below). From the consideration of the Hamiltonian matrix elements it follows that since $V_{ij} > 0$ in the expressions of bare state energies, in order to achieve an avoided crossing with the ground collective state having zero energy, the detunings have to be negative and the chirp rate has to be negative to provide a positive slope. 
   \begin{figure}
   \centerline{\includegraphics[width=7cm]{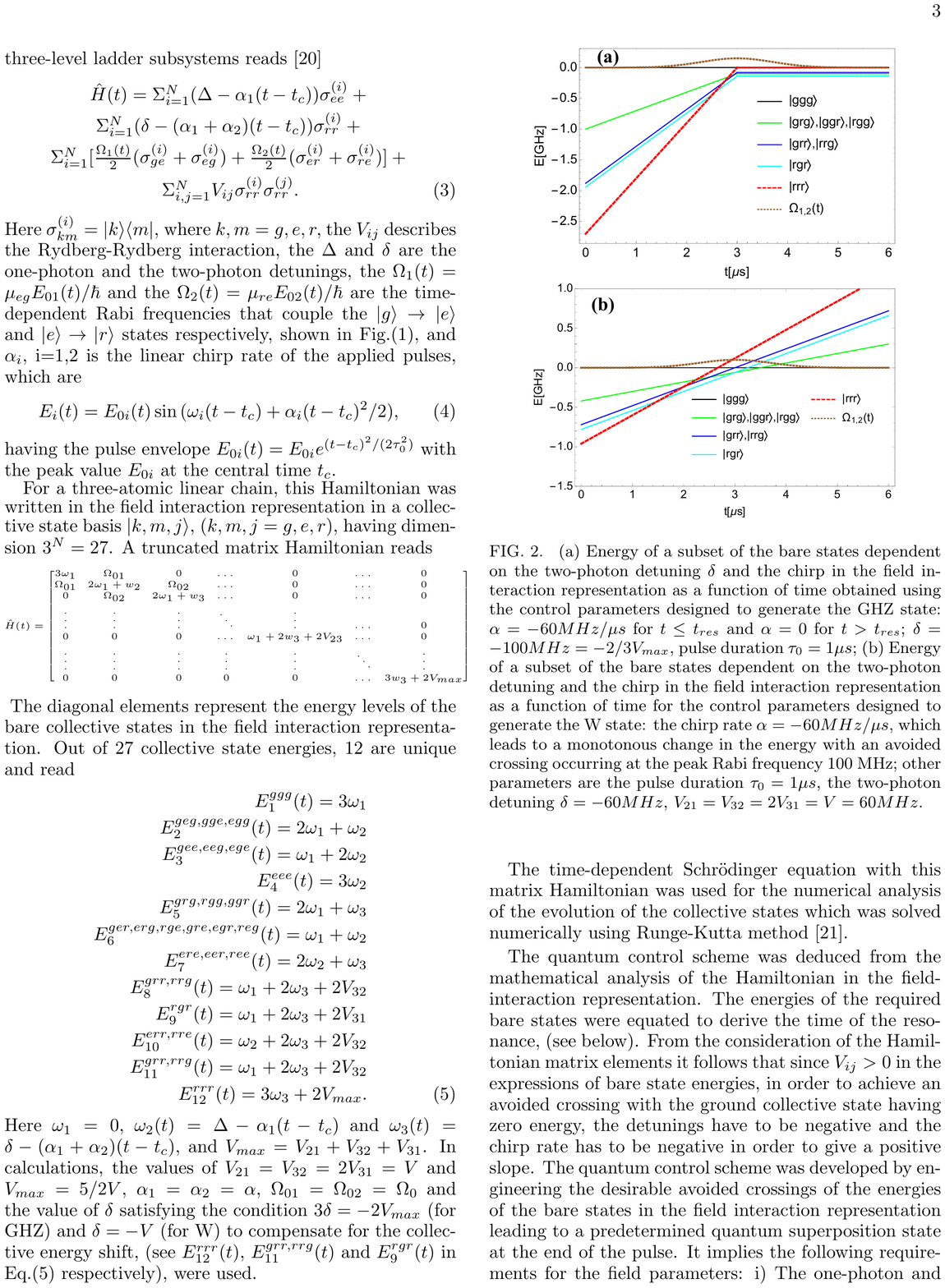}}
	\caption{\small{(a) The energies of a subset of the bare states dependent on the two-photon detuning $\delta$ and the chirp in the field interaction representation as a function of time obtained using the control parameters designed to generate the GHZ state:  $\alpha$ = -150 MHz/$\mu s$ for $t \le t_{res}$ and $\alpha = 0$ for $t > t_{res}$; $\delta$= -100 MHz=-2/3 $V_{max}$, $V_{21}=V_{32}=2V_{31}=V=60MHz$,  pulse duration $\tau_0= 1 \mu s$;  
	(b) The energies of a subset of the bare states dependent on the two-photon detuning and the chirp in the field interaction representation as a function of time for the control parameters designed to generate the W state:  $\alpha$= -60 MHz/$\mu s$, $\tau_0= 1 \mu s$,   $\delta= -60 MHz$, $V_{21}=V_{32}=2V_{31}=V=60MHz$.} 
	}
 	\label{fig2_2}
 \end{figure}
The quantum control scheme was developed by engineering the desirable avoided crossings of the energies of the bare states in the field interaction representation leading to a predetermined quantum superposition state at the end of the pulse.  It implies the following requirements for the field parameters: i) The one-photon and two-photon detunings $\Delta$ and $\delta$ must be negative in sign so that they give the starting negative values of the collective state energies; ii) Chirp rate $\alpha_i$ has to be negative 
so that the energy slope is positive; 
iii) The detunings must satisfy $|\Delta | \gg | \delta| $ so that the transitional states dependent on $\Delta$ are significantly shifted and do not resonate with $|ggg\rangle$ during the pulse duration.

In order to generate the GHZ state, the collective Rydberg state $|rrr\rangle$ energy has to perform the only avoided crossing with the energy of the ground state $|ggg\rangle$ during the pulse duration. The time of the crossing of the Rydberg state with the ground state - the resonance time - is obtained from the condition $ E_{12}^{rrr} (t)=E_1^{ggg}(t).$ It is equal to $t_{res}=(\delta/2+(V_{21}+V_{32}+V_{31})/3) / \alpha +t_c$, here $t_c$ is the time of the peak value of the Gaussian pulse envelope. If $2(V_{21}+V_{32}+V_{31})=3\delta$, the resonance occurs at the peak intensity of the field.   At the resonance, the chirp of both pulses gets turned off to prevent further population transfer and achieve a superposition state with equal probability amplitudes. Thus, for the GHZ state formation the condition on the chirp rate in Eq.(\ref{pulse}) is $\alpha \ne 0$ for $t \le t_{res}$ and $\alpha = 0$ for $t > t_{res}$.  For an experimental realization in a three-atomic spin chain using field parameters presented in this paper, the chirp may be turned off faster than  
$\sim 0.1 \mu s$ to avoid undesirable crossing with the energy of the adjacent double degenerate collective state $|rrg\rangle, |grr\rangle$ and to prevent high harmonic generation.   The energies of a subset of the bare states having avoided crossings close to $|ggg\rangle$ state and dependent on the two-photon detuning are shown in Fig.(\ref{fig2_2},(a)) as a function of time, (the states dependent on the one-photon detuning are evolving essentially below due to a large value of the $\Delta$ and, thus, are selected out from the dynamics and the figure). The magnitude of the chirp $\alpha$ is chosen such that the effective two-photon detuning is zero. Then, the energies of states change from a slope to a horizontal lines at the peak Rabi frequency when chirp is turned to zero to provide a half of the population transfer from the ground to the collective Rydberg state and to generate the three-atomic GHZ state. The dynamics of population transfer occurs essentially within a single dressed state consisting of the $|ggg\rangle$ and the $|rrr\rangle$ state. This is due to the fastest dynamics of the $|rrr\rangle$ state leading to the first avoided crossing. Since at the time of avoided crossing, $t_{res} = t_c$, the chirp is turned off, equal population distribution is achieved.

We demonstrate adiabatic passage  from $|ggg\rangle$, having energy $E_1^{ggg}$ to the $|rrr\rangle$, having energy $E_{12}^{rrr}$ leading to the creation of the GHZ state on an approximate analytical solution of the two-level system. All transitional states with  energies from $E_2^{kmj}$ to $E_{11}^{kmj}$ were adiabatically eliminated assuming their contribution to dynamics may be neglected according to the time-dependence of their energies shown in Fig.(\ref{fig2_2},(a)). Then we obtain an effective two-level system described by the field-interaction Hamiltonian with the Rabi frequency $\Omega_{eff}(t)$, which read
\begin{eqnarray} \label{effHam1}
&\dot{a}_{ggg}= i \Omega_{eff}(t) a_{rrr}  \\
&\dot{a}_{rrr}=i 6 \alpha(t-t_c) a_{rrr} + i \Omega_{eff}(t) a_{ggg} \nonumber \\
&\centerline{} \nonumber \\
& \hat{H}_{eff}=  \hbar\left(\begin{array}{ccccccc}
0 & -\Omega_{eff}(t) \\
-\Omega_{eff}(t)  & -6 \alpha (t-t_c)
\end{array}\right). \label{effHam} 
\end{eqnarray} 
 Here the effective Rabi frequency is $\Omega_{eff}(t) \sim \Omega_0^6/(\Delta^2 V^3)$. Numerical calculation of populations of $|ggg\rangle$ and $|rrr\rangle$ using Hamiltonian in Eq.(\ref{effHam}), correlates with the exact solution. Within this two-level model, we demonstrate adiabatic passage leading to a coherent superposition of two states, $|ggg\rangle$ and $|rrr\rangle$,  with equal populations. We diagonalize the effective field-interaction Hamiltonian (\ref{effHam}) ${\bf \hat{H}_d}(t)={\bf T}(t) {\bf \hat{H}_{eff}}(t) {\bf T^\dagger}(t)$ and get to the dressed state basis, in which adiabatic transition from $|ggg\rangle$ to $|rrr\rangle$ takes place within a single dressed state. Here, ${\bf T}(t)$ is a unitary transformation matrix ${\bf T}(t)={\bf I} \cos \Theta(t) -i {\bf \sigma_y } \sin \Theta (t).$ The probability amplitudes of the dressed states ${\bf c_{d}}$ and the bare states ${\bf a}$ are related as ${\bf c_d}(t)={\bf T}(t) {\bf a}(t)$. Within the adiabatic approximation, population dynamics in the two-level  system occurs within a dressed state having lower energy $| \Psi (t) \rangle = \cos \Theta (t) |ggg\rangle -\sin \Theta (t) |rrr\rangle$. Note, that the global dynamic phase is omitted here. The coefficients are matrix elements of  ${\bf T}(t)$, which read
\begin{eqnarray}
 \cos \Theta (t) = \left(  \frac{1}{2} +  \frac{3 \alpha (t-t_c)}{2\sqrt{\Omega^2_{eff}(t)+ (3 \alpha (t - t_c))^2}} \right)^{1/2} \\
 \sin \Theta (t) = \left(  \frac{1}{2} -  \frac{3 \alpha (t-t_c)}{2\sqrt{\Omega^2_{eff}(t)+ (3 \alpha (t - t_c))^2}} \right)^{1/2}
\end{eqnarray}
The control scheme, which implies  $\alpha \le 0$ for $t \le t_c$ and  $\alpha = 0$ for $t > t_c$, provides at t=0 the value of $\cos \Theta (0)=1,$ and $\sin \Theta (0)=0,$ meaning that the population is initially in the ground $|ggg\rangle$ state. Then at $t=t_c$, probability amplitudes of two states change to become the same in magnitude: $ \cos \Theta (t_c)=\frac{1}{\sqrt{2}}$ and $ \sin \Theta (t_c) =  \frac{1}{\sqrt{2}}$. At later times, $t > t_c$, the chirp is set to zero, and the field is in the resonance with the transition frequency of the system, therefore any changes of the population are suppressed (the applied field changes only the global dynamical phase), preserving created superposition state. Thus, at the end of the pulse the probability amplitudes of $|ggg\rangle$ and $|rrr\rangle$ states are $ \cos \Theta (t_{\infty})=\frac{1}{\sqrt{2}}$ and $ \sin \Theta (t_{\infty}) =  \frac{1}{\sqrt{2}}$. 
If to maintain the constant chirp rate till the end of the pulse duration such that the same $\alpha \le 0$ is applied for $t \le t_c$ and $t > t_c$, population transfers 100\% from the ground to the excited state because $ \cos \Theta (t_{\infty})=0$ and $ \sin \Theta (t_{\infty}) =  1$. Such a control approach was implemented in the case of the W state creation. Specifically, two applied pulses were chirped with the same constant chirp during the whole pulse duration, which resulted in the population transfer from the ground $|ggg\rangle$ state to an effective excited state, which is a superposition of $|rrg\rangle$, $|grr\rangle$, and $|rgr\rangle$ states.

However, in contrast to the GHZ state generation scheme, the W state preparation is principally non-adiabatic. A subset of states relevant for the dynamics to generate the W state and their energies as a function of time are shown in Fig.(\ref{fig2_2},(b)).  Since only the $|ggg\rangle$ state is populated initially, and the target $|W\rangle$ state, Eq.~(\ref{Wdefine}), is the equal superposition of three bare states, $|rrg\rangle, |grr\rangle$, and $|rgr\rangle$, at least two more dressed stated have to be involved to achieve the goal of reaching the $|W\rangle$ state at the final time. The initially populated dressed state correlates with the $|ggg\rangle$ state at t=0. However, this dressed state correlates with $|rrr\rangle$ state at final time, thus it must be empty at the end of the pulse. Therefore, two other dressed states must be involved, which correlate with $|grr\rangle$, $ |rrg\rangle$ and $|rgr\rangle$ states at the final time. To achieve the target $|W\rangle$ state, the crossing between $|ggg\rangle$ and $ |rrr\rangle$ states must be diabatic, meaning that no population is transferred from the $|ggg\rangle$ state to the $|rrr\rangle$ state. Next crossing between the $|ggg\rangle$  and the degenerate $|grr\rangle$ and $|rrg\rangle$ states must be partially adiabatic, such that 2/3 of the population goes to those states and 1/3  stays in the $|ggg\rangle$ state.  This 1/3 of the population goes to the $|rgr\rangle$  state at the next crossing  between $|ggg\rangle$ and $|rgr\rangle$  state.  Note, that the time evolution of the energies of the states shown in Fig.(\ref{fig2_2},(a),(b)) for the strength of Rydberg-Rydberg interactions $V_{21}=V_{32}=V$ and  $V_{31}=V/2^6 $ is qualitatively the same. 

\begin{figure}  \label{fig2}
	\centerline{
		\includegraphics[width=11.cm]{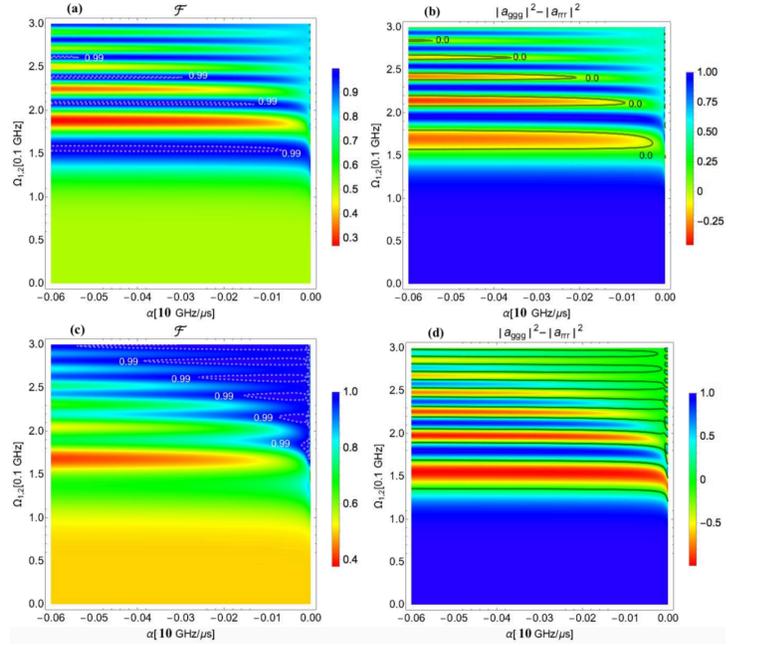}}  
	\caption{ \small{
		 (a) Fidelity of the GHZ state as a function of the chirp rate and the peak Rabi frequency for $V_{31}=V/2$,    $V_{max}= 150 MHz$; (b) The population difference of states $|ggg\rangle$ and $|rrr\rangle$ as a function of the chirp rate and the peak Rabi frequency for $V_{31}=V/2$, $V_{max}= 150 MHz$. (c) Fidelity of the GHZ state as a function of the chirp rate and the peak Rabi frequency for $V_{31}=V/2^6 $, $V_{max}= 120.94 MHz$; (d) The population difference of states $|ggg\rangle$ and $|rrr\rangle$ as a function of the chirp rate and the peak Rabi frequency for $V_{31}=V/2^6 $, $V_{max}= 120.94 MHz$. Parameters used in calculation are $V=60 MHz$, $\tau_0 = 1\mu s$, $\Delta=-1.5 GHz$, and    $\delta =-2/3 V_{max}.$  }}
 \end{figure}

The GHZ and the W state generation  is demonstrated below using numerical solution of the Sch\"odinger equation.  In calculations to generate the GHZ state the following values of the parameters of the fields were used: the pulse duration $\tau_0 = 1\mu s$, the one-photon detuning $\Delta=-1.5$ GHz, and the peak Rabi frequency of both applied pulses $\Omega_{01(2)}$  in the range from 0 to 300 MHz,  the chirp rate $\alpha_{1,2}$ in the range from 0 to -600 MHz / $\mu s$, and the two-photon detuning  $\delta =-2/3 V_{max}$, which has two values, $\delta=-100 MHz$ for $V_{31}=V/2$ and $\delta=- 80.63 MHz$ for $V_{31}=V/2^6 $  for nearest neighbor interaction $V=60MHz$. Numerical analysis demonstrates that the generation of the GHZ state is possible via two-photon adiabatic passage with the same chirp rate $\alpha_{1,2}=\alpha$ on two pulses and the same overlapping Rabi frequencies $\Omega_{01(2)}(t)$. The fidelity of the GHZ state was calculated as
\begin{equation}
F=1/2(\mid a_{ggg} \mid^2+\mid a_{rrr} \mid^2+2 Re(a_{ggg} a_{rrr}^{\dagger})).
\end{equation}
The fidelity of the GHZ state is shown in Figs.(\ref{fig2},(a),(c)) as a function of the chirp rate and the strength of the peak Rabi frequency for  $V_{31}=V/2$ (a) and $V_{31}=V/2^6 $ (c). The optimal values of the fidelity within 0.995 range are highlighted by the contour plots. Figs.(\ref{fig2},(b),(d)) show the density plot of the difference between the populations of the $|ggg\rangle$ and the $|rrr\rangle$ states and brings an additional information for a complete picture of the GHZ state formation. The zero values of the population difference indicate upon equal populations of these states, while fidelity about 0.995 in the same region of the field parameters carries information about high phase correlation between atoms. 
A comparison of the results for   $V_{31}=V/2$ (a),(b)  and $V_{31}=V/2^6 $ (c),(d) suggests that the generation the GHZ state with high fidelity, a wider choice of the field parameters is possible for a smaller value of the interaction between terminal atoms $V_{31}$. For $V_{31}=V/2^6 $, the results are more robust for experimental realization, but at the expense of a higher field amplitude. 
		 
\begin{figure}
	\centerline{
		\includegraphics[width=6.5cm]{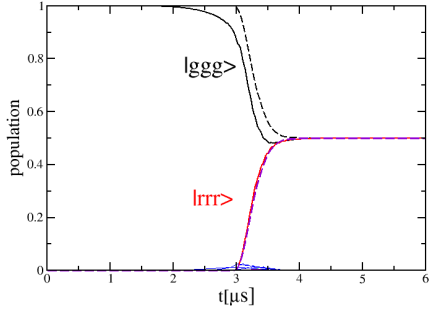}}
	\caption{ \small{The time dependence of the population of the  $|ggg\rangle$ and $|rrr\rangle$ states leading to the formation of the GHZ state at the end of the pulse duration for parameters of the field $\tau_0 = 1\mu s$, $\Delta=-1.5 GHz,$  $\delta=-100MHz,$ $\Omega_{01(2)}=158 MHz,$ $\alpha=-176 MHz/ \mu s$, and $V_{max}=150MHz.$ Dashed curves show an approximate solution using Eqs.(\ref{effHam1}). }} 
	\label{adiabatic}
\end{figure}

 The time dependence of the population of the  $|ggg\rangle$ and $|rrr\rangle$ states leading to the formation of the GHZ state at the end of the pulse duration is shown in Fig.(\ref{adiabatic}) demonstrating adiabatic passage for parameters of the field $\Delta=-1.5 GHz$, $\delta=-100MHz$, $\Omega_{01(2)}=158 MHz$, $\alpha=-176 MHz/ \mu s$, and $V_{max}=150MHz$. 	
 
\begin{figure}
	\centerline{
		\includegraphics[width=11.cm]{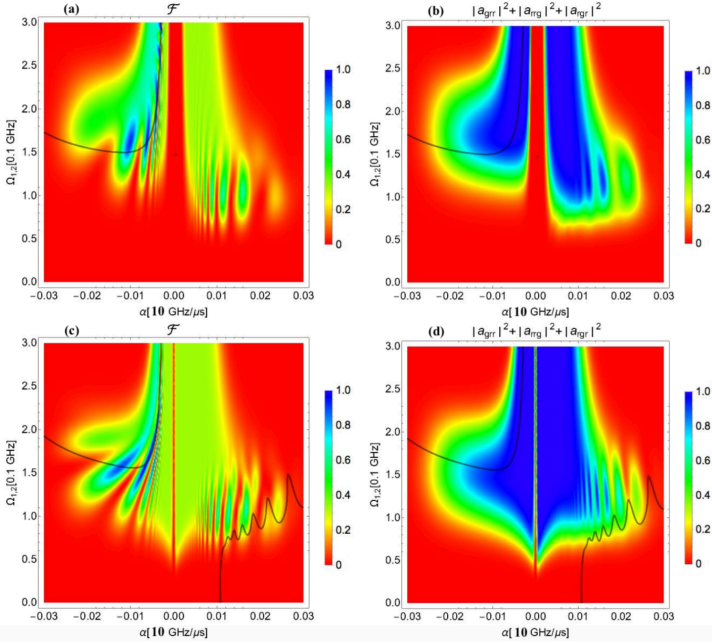}}  
	\caption{ \small{(a) Fidelity of the W state as a function of the chirp rate and the peak Rabi frequency for $V_{31}=V/2$,  $V_{max}= 150 MHz$; (b) The density plot of the sum of populations of $|rrg\rangle, |grr\rangle$, and $|rgr\rangle$ states for    $V_{31}=V/2$, $V_{max}= 150 MHz$. (c) Fidelity of the W state as a function of the chirp rate and the peak Rabi frequency for $V_{31}=V/2^6, $ $V_{max}= 120.94 MHz$; (d) The density plot of the sum of populations of $|rrg\rangle, |grr\rangle$, and $|rgr\rangle$ states for $V_{31}=V/2^6, $ $V_{max}= 120.94 MHz$. Parameters used in calculation are $V=60THz$, $\tau_0 = 1\mu s$, $\Delta=-1.4 GHz$, and   $\delta =-4.5 MHz.$ The black curves draw the contour where three contributing states have equal population.	 }}
	\label{Wstate}
\end{figure}

 In numerical calculations of the W state generation the following values of the parameters of the fields were used: The pulse duration $\tau_0 = 1\mu s$,  the chirp rate $\alpha_{1,2}$ in the range from 0 to $-300 MHz / \mu s$,  the two-photon detuning $\delta =-4.5 MHz$, the one-photon detuning $\Delta=-1.4 GHz$, and the peak Rabi frequency of both applied pulses $\Omega_{01(2)}$  in the range from 0 to $300 MHz$. Two values of the parameter $V_{31}$ were used,  $V_{31}=V/2$ and $V_{31}=V/2^6 $, which provide $V_{max}= 150 MHz$  
 and $V_{max}= 120.94 MHz$ for nearest neighbor interaction $V=60MHz$.  
  
The fidelity of the W state was calculated as
\begin{eqnarray}
&F=1/2(\mid a_{rrg} \mid^2+\mid a_{grr} \mid^2 +\mid a_{rgr} \mid^2+ \nonumber \\
&2 Re(a_{rrg} a_{grr}^{\dagger})+2 Re(a_{rrg} a_{rgr}^{\dagger})+2 Re(a_{grr} a_{rgr}^{\dagger})).
\end{eqnarray}
Figs.(\ref{Wstate},(a),(c)) show fidelity of the W state as a function of the chirp rate and the peak Rabi frequency for $V_{31}=V/2$ (a) and $V_{31}=V/2^6 $ (c). Strong dependence on $\alpha_{1,2}$ and $\Omega_{01(2)}$ is observed indicating upon nonadiabatic regime of light-matter interaction. The black curve draws the contour where three contributing states have equal population. The blue regions of fidelity through which this curve passes manifest highest values owing to equal population of the contributing states and the same phase between them. Figs.(\ref{Wstate},(b),(d)) show the density plots of the sum of populations of $|rrg\rangle, |grr\rangle$, and $|rgr\rangle$ states for $V_{31}=V/2$ (b) and $V_{31}=V/2^6 $ (d). Together with the density plots of the difference of populations between $|rrg\rangle$ and  $|rgr\rangle,$ (not shown here), they provided sufficient information to draw the contours of equal populations.

  \begin{figure}
	\centerline{
		\includegraphics[width=6.5cm]{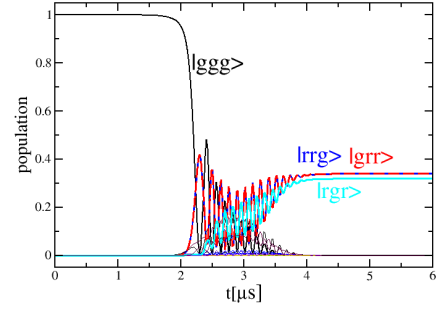}}
	\caption{ \small{Time dependence of population  of the  $|grr\rangle$, $|rrg\rangle$ and $|rgr\rangle$ states leading to the formation of the W state at the end of the pulse duration for parameters of the field $\tau_0 = 1\mu s$, $\Delta$=-1.4 GHz, $\delta$=-4.5 MHz, $\Omega_{01(2)}$=262 MHz, $\alpha$=-32 MHz/ $\mu s$, and $V_{21}=V_{32}=2 V_{31}=V$=60MHz. }} 
	\label{nonadiabatic}
\end{figure} 
By comparing the results for $V_{31}=V/2$ (a),(b) and $V_{31}=V/2^6 $ (c),(d) we conclude that a smaller value of the interaction between terminal atoms somewhat improves the results for the generation the W state, meaning that the target area of the field parameters providing higher fidelity states is increased. However, since the process of the W state generation is principally non-adiabatic with at least three dressed states  populated, the considered change in the value of $V_{31}$ does not substantially change the non-adiabatic coupling between the important dressed states, (one of which correlates with $|ggg\rangle$ state at t=0, while other correlate with $|grr\rangle,$ $|rrg\rangle$ and $|rgr\rangle$ at final time.)

The dynamics of the population of the $|rrg\rangle, |grr\rangle$, and $|rgr\rangle$ states, forming the W state is shown in Fig.(\ref{nonadiabatic}) for $\Delta$=-1.4 GHz, $\delta$=-4.5MHz,  $\Omega_{01(2)}$=262 MHz, $\alpha$=-32 MHz/ $\mu s$, and $V_{21}=V_{32}=2 V_{31}=V$=60MHz. 
The Rabi oscillations between the mostly populated states $|rrg\rangle, |grr\rangle$, and $|rgr\rangle$ are clearly shown at the intermediate times.  These  oscillations can be explained by the non-adiabatic coupling between states in the dressed state picture indicating upon the non-adiabatic nature of the W state formation. 
The fidelity of such state formation is 0.999 and is among the highest values in the provided numerical results. Parameters used in the time-dependent calculations of Figs.(\ref{adiabatic},\ref{nonadiabatic}) may be used to explore the experimental realizations of the GHZ and W states.

The lifetime of the Rydberg states is on the order of 100 $\mu s$, while the pulse duration used in our method is 1$\mu s$. Two orders of magnitude difference permits us to neglect the decoherence effects in the systems during control operations with the applied fields. Besides, one-photon detuning from intermediate state 1.5 GHz is about an order of magnitude larger than the natural bandwidth of these states known to be $\sim$10 ns \cite{Ga05,Fe67}. Such detuning results in a negligible population of transitional states minimizing decoherence. In principle, a one-photon excitation, which would need a photon in the ultraviolet range \cite{Be11}, may be used to excite atoms to a predetermined magnetic sublevel of the Rydberg state. However, the two-photon excitation scheme is more robust because it requires visible light; it is commonly used in Rydberg experiments with trapped alkali atoms. Besides, it offers a broader range of control parameters including the one-photon and the two-photon detuning, which bring flexibility to the control scheme to perform adiabatic passage on large atomic systems.  Strong Rydberg-Rydberg interactions provide significant collective energy shifts beneficial for controllable excitations of predetermined collective states.  
Meanwhile, a long lifetime of Rydberg states is efficient for quantum operations.  

{\bf Conclusion.}  Quantum control of multipartite entangled states generation involving coherent superpositions of ultracold Rydberg atoms is presented based on the two-photon passage on the selected state manifold using circularly polarized and linearly chirped pulses.  Selectivity of states is achieved through the choice of the one-photon detuning, the ratio of the Rabi frequency to the collective coupling strength and the chirp rate. The methodology is simple in experimental realization implying equal a.c. Stark shifts and linearly chirped pulses. It is assumed for a generation of entangled states of different classes in various configurations of atoms, e.g., spin chains and arrays. While quantum optimization algorithms \cite{Fa14,Zh18} are beneficial for large systems with vanishing spectral gaps, our analytical method is more practical for systems of tens of atoms.  \\

{\bf Acknowledgment. } Authors acknowledge support from the Office of Naval Research.

{}

\centerline{}\centerline{}\centerline{}

{\bf Additional Information}

\centerline{}\centerline{}

{\bf Competing interests}\\

The authors declare no competing interests.\\

{\bf Author contributions statements}\\

SM wrote the manuscript text and prepared figures. EP contributed to numerical calculations. All authors reviewed the manuscript.

\end{document}